\documentclass[a4paper]{article}

\usepackage{INTERSPEECH2022}
\usepackage{amsmath,graphicx,amssymb}
\usepackage{enumitem}
\usepackage{mathtools}
\usepackage{siunitx}
\usepackage{cleveref}
\Crefname{section}{Sec.}{Secs}  
\usepackage{bm}
\usepackage[acronym]{glossaries}
\usepackage{balance,cite,url}
\usepackage{comment}

\usepackage{tabularx, etoolbox, booktabs}
\usepackage{multirow} 
\newcolumntype{H}{>{\setbox0=\hbox\bgroup}c<{\egroup}@{}}  
\usepackage{subcaption}
\usepackage{color}
\usepackage{algorithm}
\usepackage{algpseudocode}

\usepackage{color}

\newcommand*{\tran}{^{\mkern-1.5mu\mathsf{T}\mkern-1mu}}

\newcommand{\graph}{\mathcal{G}}
\newcommand{\svert}{\mathcal{V}}    
\newcommand{\sedge}{\mathcal{E}}    
\newcommand{\vertex}{v}

\newcommand{\scoloring}{\mathcal{B}}   

\title{Utterance-by-utterance overlap-aware neural diarization with Graph-PIT \vspace{-2mm}}
\name{Keisuke Kinoshita$^1$, Thilo von Neumann$^2$, \\Marc Delcroix$^1$, Christoph Boeddeker$^2$, Reinhold Haeb-Umbach$^2$ \vspace{-2mm}}
\address{
  $^1$NTT corporation, Japan \ \ \ \ \ \ \ \ \  $^2$Paderborn University, Germany}
\email{keisuke.kinoshita@ieee.org \vspace{-3mm}}

\begin{document}

\maketitle
\begin{abstract}
Recent speaker diarization studies showed that 
integration of end-to-end neural diarization (EEND) and clustering-based diarization is a promising approach for achieving state-of-the-art performance on various tasks.
Such an approach first divides an observed signal into fixed-length segments,
then performs {\it segment-level} local diarization based on an EEND module, 
and merges the segment-level results via clustering to form a final global diarization result.
The segmentation is done to limit the number of speakers in each segment 
since the current EEND cannot handle a large number of speakers.
In this paper, we argue that such an approach involving the segmentation has several issues;
for example, it inevitably faces a dilemma that larger segment sizes increase both the context available for enhancing the performance and the number of speakers for the local EEND module to handle.
To resolve such a problem,
this paper proposes a novel framework that performs diarization without segmentation. However, it can still handle challenging data containing many speakers and a significant amount of overlapping speech.
The proposed method can take an entire meeting for inference
and perform {\it utterance-by-utterance} diarization that clusters utterance activities in terms of speakers.
To this end, we leverage a neural network training scheme called Graph-PIT proposed recently for neural source separation. 
Experiments with simulated active-meeting-like data and CALLHOME data show the superiority of the proposed approach
over the conventional methods.
\end{abstract}

\section{Introduction}
\label{sec:intro}
Automatic meeting/conversation recognition and analysis (hereafter, meeting analysis) is one of the essential technologies required 
for realizing futuristic speech applications such as communication agents that can follow, respond to, and facilitate our conversation. 
As an essential central task for the meeting analysis, speaker diarization has been extensively studied \cite{DIHARD_data, AMI_data, Diarization_review}.

Current competitive diarization approaches can be categorized into three classes: 
(1) speaker embedding clustering-based approaches \cite{Diarization_review,x-vector,DIHARD_JHU,DIHARD_BUT}, 
(2) neural end-to-end diarization (EEND) approaches \cite{Fujita_IS2019,Fujita_ASRU2019,Horiguchi2020_EDA_EEND}, 
and (3) integration of the former two approaches
\cite{EEND-vector-clustering_ICASSP2021, EEND-vector-clustering_Interspeech2021,Horiguchi_ASRU2021, coria2021overlapaware, EEND-VC-iGMM_ICASSP2022}.

The speaker embedding clustering-based approaches 
first split a recording into short homogeneous segments 
and compute speaker embeddings such as x-vectors \cite{x-vector} for each segment, assuming that only one speaker 
is active in each segment.
Then, the speaker embeddings are clustered to regroup segments belonging to the same speakers 
and obtain the diarization results. 
While these methods can cope with very challenging scenarios \cite{DIHARD_JHU,DIHARD_BUT}
and work with an arbitrarily large number of speakers,
there is a clear disadvantage that they cannot explicitly handle overlapped speech.

The second category of diarization approaches, EEND, was recently developed \cite{Fujita_IS2019,Fujita_ASRU2019,Horiguchi2020_EDA_EEND} 
to specifically address the overlapped speech problem.
Similarly to the neural source separation \cite{Kolbaek2017,RSAN},
a Neural Network (NN) receives frame-level spectral features 
and directly outputs a frame-level speaker activity for each speaker, 
no matter whether the input signal contains overlapped speech or not.
While the system is simple and has started outperforming the conventional speaker embedding clustering-based approaches \cite{Fujita_ASRU2019,Horiguchi2020_EDA_EEND},
it still has difficulty in generalizing to recordings containing a large number of speakers \cite{Horiguchi2020_EDA_EEND}.

The third category of approaches combines the strength of the previous two and simultaneously addresses their problems.
It integrates the previous two approaches and thus will be referred to as EEND-vector clustering (EEND-VC) hereafter. 
It first splits the input recording into fixed-length {\it segments} to limit the number of speakers in each segment.
Then, it applies a modified version of EEND to each segment
to obtain diarization results for speakers speaking in each segment 
as well as speaker embeddings for them. 
Finally, to estimate which of the diarization results obtained in local segments
belongs to the same speaker, 
constrained speaker clustering is performed across the segments 
based on the speaker embeddings.
Note that the original EEND-VC framework \cite{EEND-vector-clustering_ICASSP2021} 
works on a strict assumption that the number of speakers in each segment is equal to or less than the number of NN output channels, which is typically 2 or 3.\footnote{\cite{Horiguchi_ASRU2021} does not have this assumption, 
but still needs the segmentation to limit the number of speakers for the segment-wise EEND module}
This integrated approach is shown to outperform the first and second approaches,
and achieves state-of-the-art results for real conversational data
with an arbitrarily large number of speakers, such as CALLHOME data \cite{EEND-vector-clustering_ICASSP2021,Horiguchi_ASRU2021, WavLM}.

Although the EEND-VC framework was found to be effective in achieving good diarization performance,
we argue that such an approach that involves the segmentation has several issues;
(1) The assumption on the number of speakers in each segment is problematic.
It immediately faces a dilemma that, if we increase the size of the segment to obtain more context to enhance the diarization performance, 
the assumption tends to be violated.
On the other hand, if we decrease the size of the segment, it deteriorates the overall diarization performance because of the lack of context.
(2) It is not suited to achieve optimal meeting analysis, which should be one of the essential goals for any diarization research.
Specifically, the segmentation inevitably split utterances into multiple pieces irrelevant to semantics,
during NN training and inference. 
It may make it hard to train such a diarization system jointly with a back-end system such as ASR,
since each segment during NN training may not always contain full utterances.
(3) The segmentation is not desirable in performing diarization, either. 
Segmenting an input recording without considering utterance boundaries can result in segments with very short fragments of utterances, which are difficult for a system to diarize.

\begin{figure*}[th]
 \begin{center}
  \includegraphics[width=118mm]{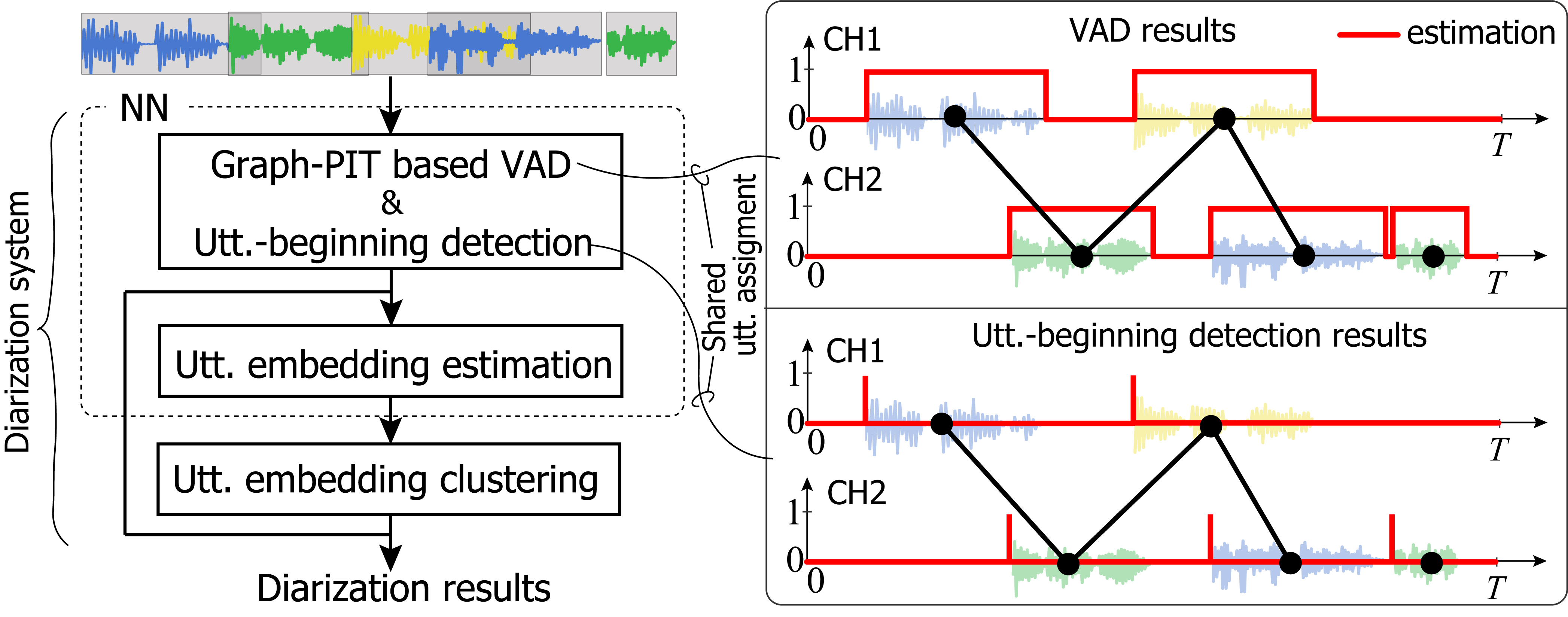}
     \end{center}
   \vspace{-3mm}
   \caption{Schematic diagram of the proposed framework. The colored waveforms depicted on the right-hand side of this figure are not directly estimated in the proposed method but shown here to clarify the relationship between each speaker's utterance and corresponding estimated VAD and utterance-beginning detection results. What is estimated by NN is shown with solid red lines. The black-colored graphs are used in Graph-PIT-based utterance assignment process explained in \ref{sec:Graph-PIT}. ``Utt.'' stands for utterance}
 \label{fig:framework}
\end{figure*}

To this end, this paper proposes to extend the EEND-VC framework such that it performs batch diarization 
without segmentation.
The proposed method explicitly models each utterance activity in a recording 
and performs {\it utterance-by-utterance} diarization that clusters utterance activities in terms of speakers.
To realize such a framework, we utilize an NN training scheme called Graph-PIT
\cite{vonneumann21_GraphPIT,vonneumann21_SpeedingUp}, 
which was recently proposed for the training of NN-based source separation.
Compared to the conventional EEND-VC framework,
the proposed method employs a much less restrictive assumption that 
the number of simultaneously-speaking speakers at any time point in a recording never exceeds the number of NN output channels.
Note that the total number of speakers in a batch processing block, which can be an entire meeting, 
can be arbitrarily large.
The relaxed assumption allows us to apply the proposed method to more diverse data, including very active casual conversations.
In the sequel, we will detail the proposed framework and experimentally show the effectiveness of the proposed method.

\section{Proposed method}
\vspace{-1mm}
\subsection{Overall framework}
\label{sec:framework}
\vspace{-1mm}
Fig.~\ref{fig:framework} shows how the proposed method performs utterance-by-utterance overlap-aware diarization. 
The schematic diagram is shown in the left half of the figure,
while its right half shows how the proposed method detects each utterance activity in an input signal.

The example observed data in Fig.~\ref{fig:framework} is meeting-like data containing 5 utterances spoken by 3 speakers as indicated by 3 different colors.
Some utterances are partially overlapped.
Importantly, in Fig.~\ref{fig:framework}, the maximum number of simultaneously-speaking speakers is two, 
and thus can be appropriately handled 
by NN having 2 output channels, as shown below.

Given the input signal, the proposed method first estimates the activity of each utterance 
by using an NN performing both Voice Activity Detection (VAD) and Utterance-Beginning Detection (UBD) for each utterance.
This process does not involve any explicit source separation.
The right half of the figure shows how the NN utilizes 2 output channels to estimate the activity of each utterance in an ``overlap-free'' manner
even though some of them are originally overlapped.
Specifically, the NN manages to assign the utterance activities to different output channels 
such that they do not overlap within each channel.
After estimating the VAD and UBD information, 
we calculate an embedding vector for each utterance.
Then, the utterance embedding vectors are clustered based on a given or an estimated number of speakers in the input signal, i.e., 3 in this example.
Finally, based on the clustering, VAD, and UBD results, 
we rearrange the estimated utterance VAD results in terms of the speakers to form the final diarization results.  
As it can be seen, the proposed method does not involve any explicit segmentation process that can potentially divide an utterance into multiple 
non-semantic pieces. 

In the following, we will first introduce the signal model 
and then detail essential components of the proposed method.

\subsection{Signal model and goal of the proposed method}
\vspace{-1mm}
The observed signal in this paper is a single-channel meeting-like data $\bar{\mathbf{X}} \in \mathbb{R}^{F \times T}$ 
in the short-time Fourier transform (STFT) domain, containing $U$ utterances $\tilde{\mathbf{S}}_u \in \mathbb{R}^{F \times T_u}\ (u=1,\ldots,U)$ uttered by $K$ speakers.
$F$, $T$, and $T_u$ stand for the total number of STFT frequency bins, the total number of time frames of the input signal,
and the total number of time frames of each utterance $\tilde{\mathbf{S}}_u$, respectively.
For convenience, we append and prepend zeros to $\tilde{\mathbf{S}}_u$ such that the resultant length matches the length of the input data, $T$,
and the resultant vector shows the temporal location of the utterance in the entire recording:
\begin{align}
\bar{\mathbf{S}}_u &= [\underbrace{\mathbf{0},\ldots,\mathbf{0}}_{T^{\textrm{(p)}}_{u}},\underbrace{
\vphantom{\mathbf{0},\ldots,\mathbf{0}}
\tilde{\mathbf{S}}_u
}_{T^{\vphantom{\textrm{(f)}}}_u},\underbrace{\mathbf{0},\ldots,\mathbf{0}}_{T^{\textrm{(f)}}_{u}}] \in \mathbb{R}^{F \times T},
\end{align}
where $\mathbf{0} \in \mathbb{R^F}$ is a zero vector.
The start timing of the $u$-th utterance in $\bar{\mathbf{X}}$ is $T^{\textrm{(p)}}_{u}+1$.
Based on these notations and a noise signal $\bar{\mathbf{N}} \in \mathbb{R}^{F \times T}$, 
the observed signal is formulated as:
\begin{align}
    \bar{\mathbf{X}} = \sum_{u=1}^U \bar{\mathbf{S}}_u + \bar{\mathbf{N}}.
\end{align}

Similarly to $\bar{\mathbf{S}}_u$, let us also define a ground-truth VAD label $\mathbf{y}_u$ and a ground-truth utterance-beginning 
label $\mathbf{z}_u$ for the $u$-th utterance as:
\begin{align}
\mathbf{y}_u &= [\underbrace{0,\ldots,0}_{T^{\textrm{(p)}}_{u}},\underbrace{1,\dots,1}_{T^{\vphantom{\textrm{(f)}}}_u},\underbrace{0,\ldots,0}_{T^{\textrm{(f)}}_{u}}]^{\tran} \in \mathbb{R}^T, \nonumber \\
\mathbf{z}_u &= [\underbrace{0,\ldots,0}_{T^{\textrm{(p)}}_{u}},1,\underbrace{0,\ldots,0}_{T-T^{\textrm{(p)}}_{u}-1}]^{\tran} \in \mathbb{R}^T. \nonumber
\end{align}
In addition, we define the feature matrix, which is the input to the NN, as 
$\mathbf{X} \in \mathbb{R}^{J \times T}$. $\mathbf{X}$ is computed from  $\bar{\mathbf{X}}$. 
In this paper, log-Mel frequency features are used.

The utterance-by-utterance diarization task we tackle here is to estimate the above VAD label associated with a (relative) speaker label for each utterance.

\subsection{Graph-PIT based overlap-aware VAD}
\label{sec:Graph-PIT}
\vspace{-1mm}
This section explains the core of the proposed framework.
As explained in \ref{sec:framework},
we can estimate the activity of each utterance in an ``overlap-free'' manner, 
if the number of the NN output channels, $C$, is equal to or outnumbers the maximum number of simultaneously-speaking speakers in the observed signal.
Let us explain how we can train an NN to perform such VAD.

With the following NN processing, we can estimate a VAD label for all $C$ output channels and all $T$ time frames,
that is, $\hat{\mathbf{Y}}=[\hat{\mathbf{y}}_1,...,\hat{\mathbf{y}}_T]^{\tran} \in \mathbf{R}^{T \times C}$ where $\hat{\mathbf{y}}_t=[\hat{y}_{t,1},\ldots,\hat{y}_{t,C}]^{\tran}$, as:
\begin{eqnarray}
    \bigl[\mathbf{h}_{1},\ldots,\mathbf{h}_{T} \bigr] &=& \mathrm{Encoder}( \mathbf{X} ) \in \mathbb{R}^{D \times T}, \nonumber \\
    \hat{y}_{t,c} &=& \mathrm{sigmoid}(\mathrm{Linear}_c^{\textrm{(VAD)}}(\mathbf{h}_{t})) \in (0,1), \nonumber \\ 
                      && \ \ \ \ \ \ \ \ \ \ \ \ \ \ \ \ \ \ \ \ \ \ \ \ \ \ \ c \in \{1,\ldots,C\}. \label{eq:vad_out}
\end{eqnarray}
$\mathrm{Encoder}(\cdot)$ is an encoder such as a multi-head self-attention NN \cite{Fujita_ASRU2019},
which utilizes all the input features $\mathbf{X}$ to obtain a frame-level internal representation $\mathbf{h}_{t}$.
$\mathrm{Linear}_c^{\textrm{(VAD)}}(\cdot) : \mathbb{R}^{D} \rightarrow \mathbb{R}^{1}$ 
is a fully-connected layer to estimate the VAD label at the output channel $c$.
$\mathrm{sigmoid}(\cdot)$ is the element-wise sigmoid function.

To ensure that the estimated VAD label $\hat{\mathbf{Y}}$ conforms to the ones in Fig.~\ref{fig:framework},
we leverage a Graph-PIT training scheme \cite{vonneumann21_GraphPIT}.
First, let us form a reference VAD label, $\mathbf{Y} \in \mathbf{R}^{T \times C}$ 
by assigning each utterance in the observation to one of the output channels.
Such an assignment can be done by the multiplication of 
an assignment matrix $\mathbf{P}=[\mathbf{p}_1, ..., \mathbf{p}_U]\tran \in \{0,1\}^{U \times C}$
and a matrix containing all the ground-truth utterance-wise VAD labels, 
$\mathbf{Y}^{\textrm{(all)}}=[\mathbf{y}_1,...,\mathbf{y}_U] \in \mathbb{R}^{T \times U}$,
i.e., $\mathbf{Y}=\mathbf{Y}^{\textrm{(all)}}\mathbf{P}$.
Each row $\mathbf{p}_u \in \mathbb{R}^C$ in $\mathbf{P}$ is a one-hot vector that indicates which output channel the $u$-th utterance is assigned to.
To make the reference VAD labels ``overlap-free'',
the assignment matrix cannot be arbitrary, but must fulfill a Graph-PIT condition which we will detail later, i.e., $\mathbf{P} \in \mathcal{B}$.
Given these notations, 
the loss function for Graph-PIT-based overlap-aware VAD is formulated as:
\begin{align}
    \mathcal{L}^{\textrm{(VAD)}} &= \min_{\mathbf{P} \in \mathcal{B}} \textrm{BCE}(\mathbf{Y}^{\textrm{(all)}}\mathbf{P},\hat{\mathbf{Y}}),
    \label{eq:graph-pit-mat}
\end{align}
where $\mathrm{BCE}(\cdot, \cdot)$ is the binary cross-entropy function between the reference and the estimated labels.

Now, let us discuss $\mathcal{B}$ that forces the assignment matrix $\mathbf{P}$
to produce always valid, i.e., ``overlap-free'', VAD labels by $\mathbf{Y}^{\textrm{(all)}}\mathbf{P}$.
As it can be guessed from Fig.~\ref{fig:framework},
the assignment matrix $\mathbf{P}$ is only valid if it does not map activities of two temporally overlapping utterances 
to the same output channel. In other words, the following constraint must be true for each assignment matrix within $\mathcal{B}$:
\begin{align}
    \mathbf{p}_u \neq \mathbf{p}_v \text{ if $u$ overlaps with $v$},\quad \forall u,v \in \{1,...,U\}.
    \label{eq:coloring}
\end{align}
The above condition is equivalent to a graph coloring problem of the unweighted and undirected overlap graph $\graph=(\svert,\sedge)$,
if we assume the graph's vertices $\svert$ correspond to utterances and its edges $\sedge$ are inserted between two vertices corresponding to 
two temporally overlapping utterances, as:
\begin{align}
    \svert &= \{1,...,U\},\\
    \sedge &= \{\{\vertex,u\}\text{ if } \vertex \text{ overlaps with }u,\quad\forall\vertex, u \in\svert\}.
\end{align}
%
An example of such a graph is drawn in the right half of Fig.~\ref{fig:framework}.
A proper coloring of $\graph$, i.e., a coloring that satisfies \cref{eq:coloring}, with $C$ colors, is then equivalent to an assignment of 
$U$ utterances to $C$ output channels.
To calculate $\mathcal{L}^{\textrm{(VAD)}}$, 
we have to find, in $\scoloring$, the one optimal coloring $\mathbf{P}^{*}$ that minimizes the term $\textrm{BCE}(\mathbf{Y}^{\textrm{(all)}}\mathbf{P},\hat{\mathbf{Y}})$ in \cref{eq:graph-pit-mat}, and then use the lowest score for back-propagation.
We can efficiently search such $\mathbf{P}^{*}$ 
by incorporating a dynamic-programming idea in \cite{vonneumann21_SpeedingUp}, 
but this part will be omitted from this paper because of the limited space.
The source code to calculate this loss is available at \cite{graph-pit-bce-code}.

\subsection{Utterance-beginning detection}
\vspace{-1mm}
Given the estimated ``overlap-free'' VAD result at each output channel,
we may be able to determine all utterance locations by examining all onsets and offsets in $\hat{\mathbf{Y}}$.
However, in the worst-case scenario,
the ground-truth activities of the $\vertex$-th and the $u$-th utterances assigned to the same channel 
can be adjacent in time, e.g., $T^{\textrm{(p)}}_{\vertex}+T_{\vertex}=T^{\textrm{(p)}}_{u}$.
In such a case, we cannot find out an utterance boundary between them
just by looking at the output ``overlap-free'' VAD result $\hat{\mathbf{Y}}$.%
\footnote{This is an issue mainly during inference, 
since during training, we always have access to the ground-truth utterance boundary labels.}
To avoid such a situation during inference,
we propose to additionally estimate beginnings of all utterances with the same NN.

Reference labels for the utterance beginning detection, $\mathbf{Z} \in \{0,1\}^{T \times C}$,
can be generated in a similar way as $\mathbf{Y}$
by using the optimal assignment matrix $\mathbf{P}^{*}$
and a matrix containing all the ground-truth utterance-wise 
utterance-beginning labels, 
$\mathbf{Z}^{\textrm{(all)}}=[\mathbf{z}_1,...,\mathbf{z}_U] \in \mathbb{R}^{T \times U}$ as 
$\mathbf{Z}=\mathbf{Z}^{\textrm{(all)}}\mathbf{P}^{*}$.
Based on this reference label, 
a loss function for the utterance-beginning detection 
is formulated as:%
\footnote{To account for the imbalance between the amount of ones and zeros in $\mathbf{Z}$,
we slightly modified $\mathbf{Z}$ as proposed in \cite{Speaker_change_detection_Interspeech2017}.}
\begin{align}
    \mathcal{L}^{\textrm{(UBD)}} &= \textrm{BCE}(\mathbf{Z}^{\textrm{(all)}}\mathbf{P}^{*},\hat{\mathbf{Z}}),
    \label{eq:UBD}
\end{align}
where $\hat{\mathbf{Z}}$ is a quantity estimated similarly to eq.~(\ref{eq:vad_out}) as:
\begin{eqnarray}
    \hat{z}_{t,c} &=& \mathrm{sigmoid}(\mathrm{Linear}_c^{\textrm{(UBD)}}(\mathbf{h}_{t})) \in (0,1). \label{eq:ubd_out}
\end{eqnarray}

\subsection{Utterance/speaker embedding estimation}
\vspace{-1mm}
To cluster the obtained utterance activities in terms of speakers, 
we also estimate an utterance/speaker embedding, $\hat{\mathbf{e}}_{u}$.
Specifically, for an utterance $u$ assigned to the $c$-th channel,
an $L$-dimensional embedding vector $\hat{\mathbf{e}}_{u}$ can be obtained as:
\begin{eqnarray}
    \hat{\mathbf{e}'}_{t,c} &=& \mathrm{Linear}_c^{\textrm{(EMB)}}(\mathbf{h}_{t}) \in \mathbb{R}^{L}, \nonumber \\
    \hat{\mathbf{e}}_{u} &=& \frac{\sum_{t=T^{\textrm{(p)}}_{u}+1}^{T^{\textrm{(p)}}_{u}+T_u} \hat{y}_{t,c} \hat{\mathbf{e}'}_{t,c}}{\|\sum_{t=T^{\textrm{(p)}}_{u}+1}^{T^{\textrm{(p)}}_{u}+T_u} \hat{y}_{t,c} \hat{\mathbf{e}'_{t,c}}\|} \in \mathbb{R}^{L}. \label{eq:speaker_embedding}
\end{eqnarray}
For the lower and upper limits for the sum in eq.~(\ref{eq:speaker_embedding}),
we use ground-truth during training,
and their estimation during inference.
The embeddings and parameters of $\mathrm{Linear}_c^{\textrm{(EMB)}}(\cdot)$ are optimized through a loss function $\mathcal{L}^{\textrm{(EMB)}}$ 
proposed in \cite{EEND-vector-clustering_ICASSP2021} as $\mathcal{L}_{\textrm{speaker}}$,
which encourages the embeddings to have small intra-speaker and large inter-speaker euclidean distances.

\subsection{Overall training loss function}
\vspace{-1mm}
NN parameters are optimized through the following multi-task loss function:
\begin{eqnarray}
    \mathcal{L} = \alpha \mathcal{L}^{\textrm{(VAD)}} + \gamma \mathcal{L}^{\textrm{(UBD)}} + \lambda \mathcal{L}^{\textrm{(EMB)}}. \label{eq:multitask_loss}
\end{eqnarray}


\section{Experiments}
\vspace{-1mm}
We evaluate the effectiveness of the proposed Graph-PIT-based utterance-by-utterance diarization (hereafter, Graph-PIT-EEND-VC)
in comparison with conventional EEND-VC,
based on simulated active meeting-like data and CALLHOME (CH) data.
As an evaluation metric,
we use diarization error rate (DER) with a collar tolerance of 0.25~s as in \cite{Horiguchi2020_EDA_EEND,Diarization_review}.

\subsection{Configurations of baselines and the proposed method}
\vspace{-1mm}
We prepared 2 versions of the conventional EEND-VC; 
The first one assumes at most 3 speakers appear within each 30-second segment (EEND-VC-30s, hereafter),
and the second one assumes at most 3 speakers appear within each 5-second segment (EEND-VC-5s, hereafter).
EEND-VC-30s achieved excellent performance on the CH data \cite{EEND-vector-clustering_Interspeech2021} 
because the assumption made therein matches the CH data closely,
but we argue that the assumption does not hold in many casual conversations, e.g., \cite{chime5}.
EEND-VC-5s employs a little more relaxed assumption, but its performance is slightly worse than the EEND-VC-30s on the CH data.
However, it is still very competitive and, for example much better than a most widely used diarization approach, i.e., x-vector clustering
as shown in \cite{EEND-VC-iGMM_ICASSP2022}.

The proposed Graph-PIT-EEND-VC tested here has 2 output channels ($C=2$) and thus assumes 
that the maximum number of simultaneously-speaking speakers are two, which is a realistic assumption 
because 3 or more people rarely speak simultaneously in meetings except for laughter segments \cite{overlap_analysis_Interspeech2006}.
For the training of the proposed method, we set the multi-task weights at $\alpha=1.0$, $\gamma=0.1$, $\lambda=0.03$.
For the clustering, we used constrained agglomerative hierarchical clustering \cite{EEND-vector-clustering_Interspeech2021} for both conventional methods and the proposed method
with the same hyper-parameter sets. 
For the proposed method, we used a cannot-link constraint between embeddings of utterances that are estimated as overlapped utterances by comparing the estimated results of the output channels.
For all methods in all experiments, we estimated the number of speakers for the clustering, as in \cite{EEND-vector-clustering_Interspeech2021}.

All the tested methods use the same NN architecture in the encoder, i.e., a multi-head self-attention NN \cite{Fujita_ASRU2019}, and the same input features, to guarantee a fair comparison.

\subsection{Data}
\vspace{-1mm}
We trained the proposed and conventional methods 
on simulated mixtures using speech from Switchboard-2, 
Switchboard Cellular, and the NIST Speaker Recognition Evaluations.
We added noise from the MUSAN corpus \cite{MUSAN}, and reverberation based on simulated room impulse responses from \cite{Ko_2017}.
We generated 2 sets of training data. The first set (6.9k~hours) consists of 1-to-3-speaker meeting-like data generated 
based on the algorithm proposed in \cite{Fujita_IS2019} with $\beta=10$. 
This training data is used to train EEND-VC-30s, as in \cite{EEND-vector-clustering_Interspeech2021}.
The second training dataset (5.5k~hours) contains more realistically active-meeting-like data with 2 to 7 speakers.
We trained EEND-VC-5s and Graph-PIT-EEND-VC based on this second dataset by using the trained EEND-VC-30s as a seed model,
because the assumption in these two methods allows them to be trained on this active meeting-like data,
while EEND-VC-30s cannot handle it appropriately because of its rigid assumption.

To thoroughly evaluate all methods, we used two evaluation datasets.
The first one is a simulated active-meeting-like data generated in the same way as the above second training dataset,
but with a different set of speakers.
The dataset includes 500 artificial test meeting-like data, whose overlap ratio varies from 5\% to 45\%, and is 26\% on average. 
The number of speakers in each meeting was randomly sampled from 2 to 7 speakers.
The average length of the meeting data was 3.3 minutes.
For the evaluation with this data, we used models trained with the aforementioned simulated data.
The second evaluation dataset is the CH dataset \cite{CALLHOME} 
that contains 500 real telephone conversation sessions containing 2 to 6 speakers. 
Because there is a mismatch between the simulated data and the CH data, 
we adapt the models trained on the simulated data to the CH data by using a part of the CH data, as proposed in \cite{Horiguchi2020_EDA_EEND}.
The overlap ratio of the CH data is 16\%, which is much lower than typical casual conversation data such as \cite{chime5}. 

\subsection{Results with simulated active meeting-like data}
\label{sec:sim_result}
\vspace{-1mm}
Table~\ref{tbl:simdata_results} shows DERs and their breakdowns, including MIssed speech (MI), False Alarm (FA), and speaker ConFusion (CF).
The proposed Graph-PIT-EEND-VC significantly outperformed both conventional methods in all metrics.

EEND-VC-30s works much worse compared to the other two,
in this active meeting-like data.
It indeed reflects its rigid assumption that at most 3 speakers 
appear within each 30-second segment.
Thanks to the relaxed assumption in EEND-VC-5s, it outperforms EEND-VC-30s,
but still does not reach the level of Graph-PIT-EEND-VC which employs the most relaxed assumption,
and does not involve any segmentation process.

\begin{table}[t]
    \centering
    \caption{Results on simulated active meeting-like data.}
    \vspace{-3mm}
    \label{tbl:simdata_results}
    \resizebox{65mm}{!}{
    \begin{tabular}{ccccc}
        \toprule
        Method                 &  DER   & MI   &   FA   & CF   \\\midrule
        EEND-VC-30s        &  19.3  & 3.6  &  1.6   & 14.1 \\
        EEND-VC-5s         &  15.8  & 2.0  &  1.0   & 12.9 \\
        Graph-PIT-EEND-VC  &  \bf{12.6} & \bf{1.2} & \bf{0.6}   & \bf{10.4} \\
      \bottomrule
    \end{tabular}
     }
\end{table}

\subsection{Results with CALLHOME data}
\vspace{-1mm}
Here, let us compare EEND-VC-5s and Graph-PIT-EEND-VC, which worked nicely on the simulated active meeting-like data.
Results are shown in Table~\ref{tbl:CH_results_estimated_number_of_speakers}. 
EEND-VC-30s, whose assumption closely matches the CH data, can achieve
DER of $12.4~\%$ on average \cite{EEND-vector-clustering_Interspeech2021}.
However, its results are omitted from the table for comparison,
since it was found in \ref{sec:sim_result} that
its unrealistic assumption does not hold in case of 
active conversations,
and thus its application area seems quite limited.
Although Graph-PIT-EEND-VC has achieved slightly better performance on average, 
it is safe to say that the performances of the two approaches on the table are comparable on the CH data.
Considering the superior performance of Graph-PIT-EEND-VC before adaptation, 
i.e., its performance on the simulated meeting-like data,
it seems to have difficulty in adapting its behavior to the CH data with a small amount of adaptation data (8.7 hours).

Overall, the obtained results are very encouraging.
Even if the performance of Graph-PIT-EEND-VC is comparable to conventional methods in some cases,
it can now perform utterance-by-utterance overlap-aware diarization which can open up many new opportunities such as joint training with ASR.
We will investigate such potentials as well as the difficulty of the adaptation in the future work.


\begin{table}[t]
    \centering
    \caption{DERs (\%) on CALLHOME data.} 
    \vspace{-3mm}
    \label{tbl:CH_results_estimated_number_of_speakers}
    \resizebox{\linewidth}{!}{
    \begin{tabular}{@{}cccccc c@{}}
        \toprule
        &\multicolumn{6}{c}{\# of speakers in a session}\\\cmidrule(l){2-6}
        Method                 &  2 & 3 & 4 & 5 & 6 & Avg.\\\midrule
        EEND-VC-5s          & \bf{7.0}  & 14.2     &  \bf{16.7}  & 31.6      & \bf{29.9}  & 13.7\\
        Graph-PIT-EEND-VC              & 7.1       & \bf{12.6} &  18.3      & \bf{31.1}  & 30.7  & \bf{13.5}\\
      \bottomrule
    \end{tabular}
    }
\end{table}

\section{Conclusions}
\vspace{-1mm}
This paper proposed an utterance-by-utterance overlap-aware diarization approach by leveraging a training scheme called Graph-PIT,
for better diarization performance and opening up new opportunities such as its joint training with ASR. 
Experiments with simulated active meeting-like data and CALLHOME data shows the superiority of the proposed framework.

\bibliographystyle{IEEEtran}

\bibliography{main}

\begin{thebibliography}{10}
\providecommand{\url}[1]{#1}
\csname url@samestyle\endcsname
\providecommand{\newblock}{\relax}
\providecommand{\bibinfo}[2]{#2}
\providecommand{\BIBentrySTDinterwordspacing}{\spaceskip=0pt\relax}
\providecommand{\BIBentryALTinterwordstretchfactor}{4}
\providecommand{\BIBentryALTinterwordspacing}{\spaceskip=\fontdimen2\font plus
\BIBentryALTinterwordstretchfactor\fontdimen3\font minus
  \fontdimen4\font\relax}
\providecommand{\BIBforeignlanguage}[2]{{%
\expandafter\ifx\csname l@#1\endcsname\relax
\typeout{** WARNING: IEEEtran.bst: No hyphenation pattern has been}%
\typeout{** loaded for the language `#1'. Using the pattern for}%
\typeout{** the default language instead.}%
\else
\language=\csname l@#1\endcsname
\fi
#2}}
\providecommand{\BIBdecl}{\relax}
\BIBdecl

\bibitem{DIHARD_data}
N.~Ryant, K.~Church, C.~Cieri, A.~Cristia, J.~Du, S.~Ganapathy, and
  M.~Liberman, \emph{First {DIHARD} Challenge Evaluation Plan}, 2018,
  {https://zenodo.org/record/1199638}.

\bibitem{AMI_data}
J.~Carletta, S.~Ashby, S.~Bourban, M.~Flynn, M.~Guillemot, T.~Hain, J.~Kadlec,
  V.~Karaiskos, W.~Kraaij, M.~Kronenthal, G.~Lathoud, M.~Lincoln, A.~Lisowska,
  I.~McCowan, W.~Post, D.~Reidsma, , and P.~Wellner, ``The {AMI} meeting
  corpus: A pre-announcement,'' in \emph{The Second International Conference on
  Machine Learning for Multimodal Interaction, ser. MLMI'05}, 2006, pp. 28--39.

\bibitem{Diarization_review}
X.~Anguera, S.~Bozonnet, N.~Evans, C.~Fredouille, G.~Friedland, and O.~Vinyals,
  ``Speaker diarization: A review of recent research,'' \emph{IEEE Transactions
  on Audio, Speech, and Language Processing}, vol.~20, no.~2, pp. 356--370, Feb
  2012.

\bibitem{x-vector}
D.~Snyder, P.~Ghahremani, D.~Povey, D.~Garcia-Romero, Y.~Carmiel, , and
  S.~Khudanpur, ``Deep neural network-based speaker embeddings for end-to-end
  speaker verification,'' in \emph{Proc. IEEE Spoken Language Technology
  Workshop}, 2016.

\bibitem{DIHARD_JHU}
\BIBentryALTinterwordspacing
G.~Sell, D.~Snyder, A.~McCree, D.~Garcia-Romero, J.~Villalba, M.~Maciejewski,
  V.~Manohar, N.~Dehak, D.~Povey, S.~Watanabe, and S.~Khudanpur, ``Diarization
  is hard: Some experiences and lessons learned for the {JHU} team in the
  inaugural {DIHARD} challenge,'' in \emph{Proc. Interspeech 2018}, 2018, pp.
  2808--2812. [Online]. Available:
  \url{http://dx.doi.org/10.21437/Interspeech.2018-1893}
\BIBentrySTDinterwordspacing

\bibitem{DIHARD_BUT}
\BIBentryALTinterwordspacing
M.~Diez, F.~Landini, L.~Burget, J.~Rohdin, A.~Silnova, K.~Zmolikova,
  O.~Novotn{\'y}, K.~Vesel{\'y}, O.~Glembek, O.~Plchot, L.~Mo{\v s}ner, and
  P.~Mat{\v e}jka, ``{BUT} system for {DIHARD} speech diarization challenge
  2018,'' in \emph{Proc. Interspeech 2018}, 2018, pp. 2798--2802. [Online].
  Available: \url{http://dx.doi.org/10.21437/Interspeech.2018-1749}
\BIBentrySTDinterwordspacing

\bibitem{Fujita_IS2019}
Y.~Fujita, N.~Kanda, S.~Horiguchi, K.~Nagamatsu, and S.~Watanabe, ``End-to-end
  neural speaker diarization with permutation-free objectives,'' in \emph{Proc.
  Interspeech 2019}, 2019, pp. 4300--4304.

\bibitem{Fujita_ASRU2019}
Y.~Fujita, N.~Kanda, S.~Horiguchi, Y.~Xue, K.~Nagamatsu, and S.~Watanabe,
  ``End-to-end neural speaker diarization with self-attention,'' in \emph{Proc.
  IEEE ASRU}, 2019, pp. 296--303.

\bibitem{Horiguchi2020_EDA_EEND}
S.~Horiguchi, Y.~Fujita, S.~Watanabe, Y.~Xue, and K.~Nagamatsu, ``End-to-end
  speaker diarization for an unknown number of speakers with encoder-decoder
  based attractors,'' 2020, arXiv:2005.09921.

\bibitem{EEND-vector-clustering_ICASSP2021}
K.~Kinoshita, M.~Delcroix, and N.~Tawara, ``Integrating end-to-end neural and
  clustering-based diarization: Getting the best of both worlds,'' in
  \emph{Proc. 2021 IEEE International Conference on Acoustics, Speech and
  Signal Processing (ICASSP)}, 2021, pp. 7198--7202.

\bibitem{EEND-vector-clustering_Interspeech2021}
------, ``Advances in integration of end-to-end neural and clustering-based
  diarization for real conversational speech,'' in \emph{Proc. Interspeech},
  2021, pp. 3565--3569.

\bibitem{Horiguchi_ASRU2021}
S.~Horiguchi, S.~Watanabe, P.~Garcia, Y.~Xue, Y.~Takashima, and Y.~Kawaguchi,
  ``Towards neural diarization for unlimited numbers of speakers using global
  and local attractors,'' 2021, arXiv:2107.01545.

\bibitem{coria2021overlapaware}
J.~M. Coria, H.~Bredin, S.~Ghannay, and S.~Rosset, ``Overlap-aware low-latency
  online speaker diarization based on end-to-end local segmentation,'' 2021,
  arXiv:2109.06483.

\bibitem{EEND-VC-iGMM_ICASSP2022}
K.~Kinoshita, M.~Delcroix, and T.~Iwata, ``Tight integration of neural- and
  clustering-based diarization through deep unfolding of infinite gaussian
  mixture model,'' in \emph{Proc. 2021 IEEE International Conference on
  Acoustics, Speech and Signal Processing (ICASSP) (To appear)}, 2022.

\bibitem{Kolbaek2017}
M.~Kolb{\ae}k, D.~Yu, Z.~Tan, and J.~Jensen, ``Multitalker speech separation
  with utterance-level permutation invariant training of deep recurrent neural
  networks,'' \emph{IEEE/ACM Transactions on Audio, Speech, and Language
  Processing}, vol.~25, no.~10, pp. 1901--1913, Oct 2017.

\bibitem{RSAN}
K.~Kinoshita, L.~Drude, M.~Delcroix, and T.~Nakatani, ``Listening to each
  speaker one by one with recurrent selective hearing networks,'' in
  \emph{Proc. 2018 IEEE International Conference on Acoustics, Speech and
  Signal Processing (ICASSP)}, April 2018, pp. 5064--5068.

\bibitem{WavLM}
\BIBentryALTinterwordspacing
S.~Chen, C.~Wang, Z.~Chen, Y.~Wu, S.~Liu, Z.~Chen, J.~Li, N.~Kanda,
  T.~Yoshioka, X.~Xiao, J.~Wu, L.~Zhou, S.~Ren, Y.~Qian, Y.~Qian, J.~Wu,
  M.~Zeng, and F.~Wei, ``Wavlm: Large-scale self-supervised pre-training for
  full stack speech processing,'' \emph{CoRR}, vol. abs/2110.13900, 2021.
  [Online]. Available: \url{https://arxiv.org/abs/2110.13900}
\BIBentrySTDinterwordspacing

\bibitem{vonneumann21_GraphPIT}
T.~von Neumann, K.~Kinoshita, C.~Boeddeker, M.~Delcroix, and R.~Haeb-Umbach,
  ``{Graph-PIT: Generalized Permutation Invariant Training for Continuous
  Separation of Arbitrary Numbers of Speakers},'' in \emph{Proc. Interspeech
  2021}, 2021, pp. 3490--3494.

\bibitem{vonneumann21_SpeedingUp}
T.~von Neumann, C.~Boeddeker, K.~Kinoshita, M.~Delcroix, and R.~Haeb-Umbach,
  ``Speeding up permutation invariant training for source separation,'' in
  \emph{Speech Communication; 14th ITG Conference}, 2021, pp. 1--5.

\bibitem{graph-pit-bce-code}
\BIBentryALTinterwordspacing
 [Online]. Available: \url{https://github.com/fgnt/graph_pit}
\BIBentrySTDinterwordspacing

\bibitem{Speaker_change_detection_Interspeech2017}
C.~B. Ruiqing~Yin, Herve~Bredin, ``Speaker change detection in broadcast tv
  using bidirectional long short-term memory networks,'' in \emph{Proc.
  Interspeech}, 2017, pp. 3827--3831.

\bibitem{chime5}
J.~Barker, S.~Watanabe, E.~Vincent, and J.~Trmal, ``The fifth ’{CHiME}’
  speech separation and recognitionchallenge: Dataset, task and baselines,'' in
  \emph{Proc. Interspeech 2018}, 2018, pp. 1561--1565.

\bibitem{overlap_analysis_Interspeech2006}
O.~Cetin and E.~Shriberg, ``Analysis of overlaps in meetings by dialog factors,
  hot spots, speakers, and collection site: Insights for automatic speech
  recognition,'' in \emph{Proc. Interspeech 2006}, 2006, pp. 293--296.

\bibitem{MUSAN}
D.~Snyder, G.~Chen, and D.~Povey, ``{MUSAN}: A music, speech, and noise
  corpus,,'' 2015, arXiv:1510.08484.

\bibitem{Ko_2017}
T.~Ko, V.~Peddinti, D.~Povey, M.~L. Seltzer, and S.~Khudanpur, ``A study on
  data augmentation of reverberant speech for robust speech recognition,'' in
  \emph{Proc. 2017 IEEE International Conference on Acoustics, Speech and
  Signal Processing (ICASSP)}, March 2017, pp. 5220--–5224.

\bibitem{CALLHOME}
M.~Przybocki and A.~Martin, \emph{{2000 NIST Speaker Recognition Evaluation
  (LDC2001S97)}}.\hskip 1em plus 0.5em minus 0.4em\relax Philadelphia, New
  Jersey: Linguistic Data Consortium, 2001.

\end{thebibliography}

\end{document}